\input{aipcheck}

\documentclass[
    ,final            % use final for the camera ready runs
  ]
  {aipproc}
\usepackage{graphicx}
\usepackage{amsfonts}
\usepackage{amssymb}
\usepackage{amsmath}
\usepackage{mathptmx}
\usepackage{ulem}
\usepackage{color}
\usepackage{shortvrb}

\layoutstyle{8x11single}

\begin{document}

\title{Lattice QCD analysis for relation between quark confinement and 
chiral symmetry breaking}

\author{Takahiro M. Doi}{
  address={Department of Physics, Graduate School of Science, Kyoto University,\\
Kitashirakawa-oiwake, Sakyo, Kyoto 606-8502, Japan}
}

\author{Hideo Suganuma}{
  address={Department of Physics, Graduate School of Science, Kyoto University,\\
Kitashirakawa-oiwake, Sakyo, Kyoto 606-8502, Japan}
}

\author{Takumi Iritani}{
  address={Yukawa Institute for Theoretical Physics, Kyoto University, \\
Kitashirakawa-Oiwake, Sakyo,  Kyoto 606-8502, Japan}
}

\begin{abstract}
The Polyakov loop and the Dirac modes are connected via a simple analytical relation 
on the temporally odd-number lattice, where the temporal lattice size is odd 
with the normal (nontwisted) periodic boundary condition. 
Using this relation, we investigate the relation between quark confinement and chiral symmetry breaking in QCD. 
In this paper, we discuss the properties of this analytical relation 
and numerically investigate each Dirac-mode contribution to the Polyakov loop 
in both confinement and deconfinement phases at the quenched level. 
This relation indicates that low-lying Dirac modes have little contribution to the Polyakov loop, 
and we numerically confirmed this fact. 
From our analysis, it is suggested that 
there is no direct one-to-one corresponding 
between quark confinement and chiral symmetry breaking in QCD. 
Also, in the confinement phase, we numerically find that there is a new ``positive/negative symmetry'' 
in the Dirac-mode matrix elements of link-variable operator which appear in the relation 
and the Polyakov loop becomes zero because of this symmetry. 
In the deconfinement phase, this symmetry is broken and the Polyakov loop is non-zero. 
\end{abstract}

\maketitle

\def\slash#1{\not\!#1}
\def\slashb#1{\not\!\!#1}
\def\slashbb#1{\not\!\!\!#1}

%%%%%%%%%%%%%%%%%%%%%%%%%%%%%%%%%%%%%%%%%%%%
%% MAINMATTER
%%%%%%%%%%%%%%%%%%%%%%%%%%%%%%%%%%%%%%%%%%%%

\section{Introduction}
One of the important problems in low-energy QCD is 
nonperturbative understanding of color confinement and chiral symmetry breaking 
and these important nonperturbative phenomena have been investigated 
in many analytical and numerical studies \cite{NJL, KS, Rothe, Greensite}. 
Their relation is also outstanding issue 
\cite{SST95, Hatta, Karsch, YAoki, Miyamura, Woloshyn, GIS, Gattringer, Langfeld, BG08, LS11}. 
Some studies suggest that confinement and chiral symmetry breaking are strongly correlated. 
In finite temperature lattice QCD calculation, some studies shows that 
the deconfinement phase transition and chiral restoration occur 
at almost the same temperatures \cite{Karsch}. 
There is an opposite study, however, that the transition temperatures of deconfinement phase transition and 
chiral restoration are not the same \cite{YAoki}. 

The order parameters of deconfinement phase transition and chiral symmetry breaking 
are important quantities to investigate the relation between confinement and chiral symmetry breaking. 
The Polyakov loop is considered as an order parameter for quark confinement \cite{Rothe}. 
At the quenched level, the Polyakov loop is the exact order parameter for quark confinement, 
and its vacuum expectation value is zero in the confinement phase 
and nonzero in the deconfinement phase. 
The order parameter of chiral symmetry breaking is chiral condensate, 
and low-lying Dirac modes are essential for chiral symmetry breaking in QCD, 
for example, according to the Banks-Casher relation \cite{BanksCasher}. 

In some lattice QCD studies, 
it is confirmed that confinement and chiral symmetry breaking are simultaneously lost 
by removing QCD monopoles in the maximally Abelian gauge \cite{Miyamura, Woloshyn}. 
These results shows QCD monopoles in the maximally Abelian gauge play important role 
for both confinement and chiral symmetry breaking. 
In previous numerical studies, however, 
it is suggested that low-lying Dirac modes are not important modes for confinement \cite{GIS}. 
In fact, the confinement properties such as the Polyakov loop and 
the string tension of quark-antiquark pair 
are almost unchanged by removing low-lying Dirac modes from QCD vacuum. 
Taking into consideration that low-lying Dirac modes are essential for chiral symmetry braking, 
this result suggests no direct one-to-one corresponding 
between confinement and chiral symmetry breaking in QCD. 

The analytical relation between the Polyakov loop and Dirac modes can be a hint for the relation between 
confinement and chiral symmetry breaking in QCD. 
In fact, for example, the Polyakov loop is expressed in terms of Dirac eigenvalues 
using twisted boundary condition for link-variables \cite{Gattringer}. 
However, 
the (anti) periodic boundary condition is required 
for the imaginary-time formalism at finite temperature. 
Recently, we derived a relation between the Polyakov loop and Dirac modes in the lattice QCD formalism 
with the normal (nontwisted) periodic boundary condition for link-variables \cite{SDI,DSIPRD,DSI}. 
First, we derived it on the temporally odd-number lattice, where the temporal lattice size is odd. 
Next, we showed that 
the analytical relation can be derived on the lattices on which the temporal lattice size is even. 
In this paper, we concentrate the temporally odd-number lattice 
because the analytical relation is especially simple form on this lattice 
and it is not serious problem to use the temporally odd-number lattice as we will discuss later. 

\section{Dirac mode and operator formalism on lattice}
In this section, we prepare the setup for this study 
and review Dirac modes and the operator formalism in SU($N_{\rm c}$) lattice gauge theory. 
We consider a standard square lattice with lattice spacing $a$ and 
sites on the lattice are denoted as $s=(s_1, s_2, s_3, s_4) \ (s_\mu=1,2,\cdots,N_\mu)$ with odd $N_{\rm 4}$. 
The temporal periodic boundary condition is imposed for link-variables $U_\mu(s)={\rm e}^{iagA_\mu(s)}$ 
with gauge fields $A_\mu(s) \in su(N_c)$, gauge coupling $g$ and sites $s=(s_1, s_2, s_3, s_4)$. 
As we will show later, the analytical relation between Polyakov loop and the Dirac modes 
can be derived for arbitrary gauge group. In this paper, however, we take SU($N_{\rm c}$) as the gauge group. 
In this paper, we define all the $\gamma$-matrices to be hermite as $\gamma^\dagger_\mu=\gamma_\mu$. 

\subsection{Dirac mode in lattice QCD}
Next, we review the Dirac mode in lattice QCD, 
which is eigenmode of the Dirac operator. 
Using link-variables $U_\mu(s)={\rm e}^{iagA_\mu(s)}$, 
the Dirac operator $\slashb{D}=\gamma_\mu D_\mu$ is expressed as 
\begin{align}
 \slashb{D}_{s,s'} 
      = \frac{1}{2a} \sum_{\mu=1}^4 \gamma_\mu 
\left[ U_\mu(s) \delta_{s+\hat{\mu},s'}
        - U_{-\mu}(s) \delta_{s-\hat{\mu},s'} \right], \label{DiracOpExp}
\end{align}
where $\hat{\mu}$ is the unit vector in direction $\mu$ in the lattice unit 
and $U_{-\mu}(s)\equiv U^\dagger_\mu(s-\hat\mu)$. 
Since $\gamma_\mu$ is Hermite in this definition, 
the Dirac operator is anti-Hermite and has pure imaginary eigenvalue, and 
the Dirac eigenvalue equation is expressed as 
\begin{eqnarray}
\slashb{D}|n\rangle =i\lambda_n|n \rangle. 
\end{eqnarray}
$i\lambda_n$ ($\lambda_n \in {\bf R}$) are the Dirac eigenvalues, 
and the Dirac eigenstates $|n \rangle$ have the orthonormality and the completeness as 
\begin{eqnarray}
\langle n|m\rangle=\delta_{mn}, \ \ \ \sum_n|n\rangle \langle n| =1. 
\end{eqnarray}
Because of $\{\slashb{D},\gamma_5\}=0$,
the chiral partner $\gamma_5|n\rangle$ is also 
an eigenstate with the eigenvalue $-i\lambda_n$.
The Dirac eigenfunction $\psi_n(s)\equiv\langle s|n \rangle $
can be determined in lattice QCD calculation except for a phase factor 
from the explicit form for the Dirac eigenvalue equation 
\begin{eqnarray}
 \frac{1}{2a}\sum_{\mu=1}^4 \gamma_\mu
[U_\mu(s)\psi_n(s+\hat \mu)-U_{-\mu}(s)\psi_n(s-\hat \mu)] =i\lambda_n \psi_n(s). \label{DiracEigenExp}
\end{eqnarray}
Because the gauge transformation of link-variables is $U_\mu(s)\rightarrow V(s)U_\mu(s)V^\dagger(s+\hat{\mu})$, 
the gauge transformation of $\psi_n(s)$ is 
\begin{eqnarray}
\psi_n(s)\rightarrow V(s)\psi_n(s), \label{GaugeTrans}
\end{eqnarray}
which is the same as that of the quark field, although 
there can appear an irrelevant $n$-dependent 
global phase factor $\mathrm{e}^{i\varphi_n[V]}$, 
according to arbitrariness of the phase in the basis $|n\rangle$ \cite{GIS}. 

Most of contribution to chiral condensate 
is given by low-lying Dirac modes, for example 
according to the Banks-Casher relation \cite{BanksCasher}: 
\begin{align}
\langle\bar{q}q\rangle=-\lim_{m\to0}\lim_{V_{\rm phys}\to\infty}\pi\langle\rho(0)\rangle, \label{BCrel} 
\end{align}
where the Dirac eigenvalue density $\rho(\lambda)$ is defined by 
\begin{align}
\rho(\lambda)\equiv \frac{1}{V_{\rm phys}}\sum_n\langle\delta(\lambda-\lambda_n)\rangle
\label{BCrel} 
\end{align}
with the space-time volume $V_{\rm phys}$. 
From Eq. (\ref{BCrel}), the chiral condensate is proportional to the Dirac zero-eigenvalue density. 
Since the chiral condensate is the order parameter 
of chiral symmetry breaking, low-lying Dirac modes are essential 
for chiral symmetry breaking. 
In general, instead of $\slashb{D}$, one can consider any (anti)hermitian operator, e.g., 
$D^2 =D_\mu D_\mu$, and the expansion in terms of its eigen-modes \cite{BI05}. 
To investigate chiral symmetry breaking, however, it is appropriate to consider $\slashb{D}$ 
and the expansion by its eigenmodes. 

The role of the low-lying Dirac modes has been studied 
in the context of chiral symmetry breaking in QCD. 
In particular, the removal of low-lying Dirac modes has been recently 
investigated to realize 
the world of ``unbreaking chiral-symmetry'' \cite{GIS, LS11}. 
For example, propagators and masses of hadrons are investigated 
after the removal of low-lying Dirac modes, 
and parity-doubling ``hadrons'' can be actually observed as bound states 
in the chiral unbroken world \cite{LS11}. 
Also, after the removal of low-lying Dirac modes from the QCD vacuum, 
the confinement properties such as the string tension are found to be 
almost kept, while the chiral condensate is largely decreased \cite{GIS}. 

\subsection{operator formalism in lattice QCD}
Next, we review the operator formalism in lattice QCD. 
We define the link-variable operator $\hat{U}_{\pm\mu}$ 
by the matrix element,
\begin{align}
\langle 
s | \hat{U}_{\pm\mu} |s' \rangle=U_{\pm\mu}(s)\delta_{s\pm\hat{\mu},s'}. \label{LinkOp}
\end{align}
Using the link-variable operator, 
the Polyakov loop $L_P$ is expressed as 
\begin{align}
L_P
=\frac{1}{N_{\rm c}V}{\rm Tr}_c \{\hat U_4^{N_4}\} 
=\frac{1}{N_{\rm c}V} \sum_s {\rm tr}_c
\{\prod_{i=0}^{N_4-1}U_4(s+i\hat{4})\}, \label{PolyakovOp} 
\end{align}
with the 4D lattice volume $V=N_1N_2N_3N_4$. 
Here, ``Tr$_c$'' denotes the functional trace of 
${\rm Tr}_c \equiv \sum_s {\rm tr}_c$
with the trace ${\rm tr}_c$ over color index. 

Using the link-variable operator, 
covariant derivative operator $\hat{D}_\mu$ 
on the lattice is expressed as 
\begin{align}
\hat{D}_\mu= \frac{1}{2a}(\hat{U}_{\mu}-\hat{U}_{-\mu}). \label{CovariantOp}
\end{align}
Thus the Dirac operator $\hat{\slashb{D}}$ on the lattice is expressed as 
\begin{align}
\hat{\slashb{D}}=\gamma_\mu \hat{D}_\mu= 
\frac{1}{2a}\sum_{\mu=1}^{4}\gamma_\mu (\hat{U}_{\mu}-\hat{U}_{-\mu}). \label{DiracOp}
\end{align}

Here, we introduce the Dirac-mode matrix element of the link-variable operator $\hat{U}_\mu$ expressed with $\psi_n(s)$: 
\begin{align}
\langle m|\hat{U}_\mu| n \rangle  
=\sum_s \langle m|s \rangle \langle s|\hat{U}_\mu| s+\hat{\mu} \rangle \langle s+\hat{\mu}|n \rangle %\nonumber \\
=\sum_s \psi_m^\dagger(s)U_\mu(s)\psi_n(s+\hat{\mu}). \label{MatEleExp}
\end{align}
Note that the matrix element is gauge invariant, apart from an irrelevant phase factor. 
Actually, using the gauge transformation Eq. (\ref{GaugeTrans}), we find the gauge transformation of
the matrix element as \cite{GIS}
\begin{align}
\langle m|\hat{U}_\mu| n \rangle 
&=\sum_s \psi_m^\dagger(s)U_\mu(s)\psi_n(s+\hat{\mu}) \nonumber \\ 
&\rightarrow
\sum_s \psi_m^\dagger(s)V^\dagger(s)\cdot V(s)U_\mu(s)V^\dagger(s+\hat{\mu})\cdot V(s+\hat{\mu})\psi_n(s+\hat{\mu}) %\nonumber \\ 
%&=\sum_s \psi_m^\dagger(s)U_\mu(s)\psi_n(s+\hat{\mu})
=\langle m|\hat{U}_\mu| n \rangle. 
\end{align}
However, the $n$-dependent global phase factor cancels as 
$\mathrm{e}^{i\varphi_n[V]}\mathrm{e}^{-i\varphi_n[V]}=1$ 
between $|n \rangle$ and $\langle n|$, and does not appear
for physical quantities such as the Wilson loop and the Polyakov loop \cite{GIS}. 
In this study, the phase factor also cancels in the diagonal matrix element $\langle n|\hat{U}_\mu| n\rangle$, 
which appears in the analytical relation shown in the next section. 

Note also that a functional trace of a product of the link-variable operators 
corresponding to the non-closed path is exactly zero 
because of the definition of 
the link-variable operator Eq. (\ref{LinkOp}): 
\begin{align}
&{\rm Tr}_c(\hat{U}_{\mu_1}\hat{U}_{\mu_2}\cdots\hat{U}_{\mu_{N}})
=
{\rm tr}_c\sum_s\langle s|\hat{U}_{\mu_1}\hat{U}_{\mu_2}\cdots\hat{U}_{\mu_{N}}|s\rangle \nonumber \\
=&{\rm tr}_c\sum_s U_{\mu_1}(s)U_{\mu_2}(s+\hat{\mu}_1)\cdots U_{\mu_{N}}(s+\sum_{k=1}^{N-1}\hat{\mu}_k)
\langle s+\sum_{k=1}^{N}\hat{\mu}_k |s\rangle =0%\nonumber \\
%=&0
\label{nonclosed}
\end{align}
with $\sum_{k=1}^{N}\hat{\mu}_k\neq 0$, 
which means the non-closed path with the length $N$. 
This is easily understood from Elitzur's theorem \cite{Elitzur} that 
the vacuum expectation values of gauge-variant operators are zero. 
Note that one cannot make any closed loop 
such as Wilson loop using odd-number link-variables on the square lattice. 
Therefore, the functional trace corresponding to the trajectories whose lattice size is odd 
cannot be gauge-invariant and is zero except for the Polyakov loop 
using the temporal periodic boundary condition. 
This fact will be used in the derivation of the analytical relation 
between the Polyakov loop and chiral symmetry breaking in the next section. 

\section{The relation between Polyakov loop and Dirac modes on the temporally odd-number lattice}
In this section, we derive the analytical relation between the Polyakov loop and the Dirac modes 
on the temporally odd-number lattice, where the temporal lattice size $N_4$ is odd 
with the temporal nontwisted periodic boundary condition for link-variables \cite{SDI,DSIPRD,DSI}. 
Then, we discuss the properties of the analytical relation 
and the relation between confinement and chiral symmetry breaking in QCD. 
The spatial lattice sizes $N_i \ (i=1,2,3)$ are taken larger than the temporal lattice size $N_4$ ($N_i >N_4$). 

\subsection{Derivation} 
\begin{figure}[h]
%\begin{center}
\includegraphics[scale=0.5]{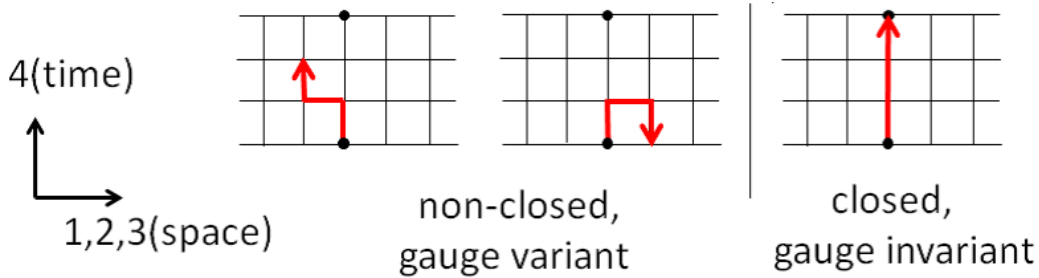}
\caption{
Some examples of trajectories corresponding to each term 
in $\hat{U}_4\hat{\slashb{D}}^{N_4-1}$ in Eq. (\ref{I}) 
on the temporally odd-number lattice with $N_4=3$. 
The left two trajectories are not closed lines and correspond to gauge variant terms, 
and the most right trajectory is a closed loop and corresponds to gauge invariant term, 
which is proportional to the Polyakov loop.
}
\label{OddLattice}
%\end{center}
\end{figure}

A key quantity for the derivation of the relation between the Polyakov loop and Dirac modes is 
\begin{eqnarray}
I\equiv {\rm Tr}_{c,\gamma} (\hat{U}_4\hat{\slashb{D}}^{N_4-1}), \label{I}
\end{eqnarray} 
where the functional trace ${\rm Tr}_{c,\gamma}$ is defined as 
%\begin{eqnarray}
${\rm Tr}_{c,\gamma}\equiv \sum_s {\rm tr}_c {\rm tr}_\gamma$
%\end{eqnarray}
including also the trace ${\rm tr}_\gamma$ over spinor index. 
Since the Dirac operator $\hat{\slashb{D}}$ is linear in the link-variable operators $\hat{U}_{\pm\mu}$, 
the quantity $\hat{U}_4\hat{\slashb{D}}^{N_4-1}$ can be expanded, 
and each term is expressed in terms of products of $N_4$ link-variable operators 
and corresponds to various trajectories whose length is $N_4$ and odd. 
In Fig. \ref{OddLattice}, some examples of the trajectories in the $N_4=3$ lattice are shown. 
Only the Polyakov loop $L_P$ and the anti-Polyakov loop $L_P^\dagger$ remain in the quantity $I$ 
because each term in the expansion of in $\hat{U}_4\hat{\slashb{D}}^{N_4-1}$ 
corresponds to the trajectories with odd length and is gauge-variant 
from the discussion around Eq. (\ref{nonclosed}) at the last in the previous section. 
Because of one temporal link-variable operator $\hat{U}_4$ in the quantity $\hat{U}_4\hat{\slashb{D}}^{N_4-1}$, 
the anti-Polyakov loop cannot be made and 
the quantity $I$ is proportional to the Polyakov loop: $I\propto L_P$. 

Actually, including the factor of proportionality, we can mathematically derive as 
\begin{align}
I
&={\rm Tr}_{c,\gamma} (\hat{U}_4 \hat{\slashb{D}}^{N_4-1}) 
={\rm Tr}_{c,\gamma} \{\hat{U}_4 (\gamma_4 \hat{D}_4)^{N_4-1}\} 
=4 {\rm Tr}_{c} (\hat{U}_4 \hat{D}_4^{N_4-1}) \nonumber \\
&=\frac{4}{(2a)^{N_4-1}}{\rm Tr}_{c} \{\hat{U}_4 (\hat{U}_4-\hat{U}_{-4})^{N_4-1}\} 
=\frac{4}{(2a)^{N_4-1}}{\rm Tr}_{c} \{ \hat{U}_4^{N_4} \} 
=\frac{4N_{\rm c}V}{(2a)^{N_4-1}}L_P. \label{I1}
\end{align}

On the other hand, 
we derive the different form of the quantity $I$ from Eq. (\ref{I1}). 
Since the Dirac mode $|n\rangle$ is complete set, 
taking Dirac modes as the basis for the functional trace in Eq. (\ref{I}), 
we find 
\begin{align}
I
=\sum_n\langle n|\hat{U}_4\slashb{\hat{D}}^{N_4-1}|n\rangle 
=i^{N_4-1}\sum_n\lambda_n^{N_4-1}\langle n|\hat{U}_4| n \rangle.  \label{I2}
\end{align}

Combining Eqs. (\ref{I1}) and (\ref{I2}), we obtain a relation between 
the Polyakov loop $L_P$ 
and the Dirac modes: 
\begin{eqnarray}
L_P=\frac{(2ai)^{N_4-1}}{4N_{\rm c}V}
%\left\langle
\sum_n\lambda_n^{N_4-1}\langle n|\hat{U}_4| n \rangle
%\right\rangle
.  \label{RelOrig}
\end{eqnarray}
This is an analytical relation between the Polyakov loop and the Dirac modes, 
i.e., a Dirac spectral representation of the Polyakov loop \cite{SDI,DSIPRD,DSI}. 

\subsection{Discussion on the analytical relation} 
Next, we discuss the properties of the analytical relation (\ref{RelOrig}) between the Polyakov loop and 
Dirac modes and analytically consider the relation between quark confinement and chiral symmetry breaking. 

\subsubsection{Properties of the analytical relation} 
Since the Polyakov loop is gauge invariant and Dirac modes can be obtained gauge-covariantly, 
this relation is gauge invariant. 
Note that, for quantitative discussion, 
we can numerically calculate each term $\frac{(2ai)^{N_4-1}}{4N_{\rm c}V}
\lambda_n^{N_4-1}\langle n|\hat{U}_4| n \rangle$ in the relation (\ref{RelOrig}) 
and investigate each Dirac-mode contribution to the Polyakov loop individually. 
Using Dirac eigenfunction $\psi_n(s)$, 
Dirac matrix element $\langle n|\hat{U}_\mu| m \rangle$ is explicitly expressed as Eq.(\ref{MatEleExp}). 
Thus, the relation (\ref{RelOrig}) is expressed as 
\begin{align}
 L_P =\frac{(2ai)^{N_4-1}}{4N_{\rm c}V}
\sum_n\lambda_n^{N_4-1}
\sum_{s}\psi^\dagger_n(s)
U_4(s) \psi_n(s+\hat{4}). 
\label{RelOrigExp}
\end{align}
Dirac eigenvalues $\lambda_n$ and Dirac eigenfunctions $\psi_n(s)$ in Eq. (\ref{RelOrigExp}) 
can be obtained 
by solving the Dirac eigenequation (\ref{DiracEigenExp}) using link-variables in each gauge configuration 
generated in Monte Carlo simulation. 

In the derivation of the relation (\ref{RelOrig}), 
we use only the following assumptions: 
\begin{enumerate}
 \item odd temporal lattice size $N_4$
 \item square lattice
 \item temporal periodicity for link-variables
\end{enumerate}
On the first assumption, it is in principle no problem to use the odd temporal size 
because the parity of the temporal lattice size does not affect the physical results in the continuum limit 
$a\rightarrow 0$ and $N_4\rightarrow \infty$. 
In fact, by a similar manner on Eq.(\ref{RelOrig}), we can also derive 
a relation which connects the Polyakov loop and Dirac modes on the 
even lattice, where all the lattice sizes are even number \cite{DSIPRD}. 
On the second one, 
physical results in the continuum limit do not depend on the lattice regularization  scheme, 
and thus it is not a problem to use the standard square lattice. 
On the third one, 
the temporal periodic boundary condition is essential for the imaginary-time finite-temperature formalism. 
In this way, the analytical relation can be derived 
without any unnatural assumption. 

The relation (\ref{RelOrig}) is satisfied for each gauge configuration, for example 
generated in the Monte Carlo simulation. 
Thus, the relation is satisfied for the gauge-configuration average: 
\begin{eqnarray}
\langle L_P\rangle=\frac{(2ai)^{N_4-1}}{4N_{\rm c}V}
\left\langle
\sum_n\lambda_n^{N_4-1}\langle n|\hat{U}_4| n \rangle
\right\rangle.  \label{RelOrigVEV}
\end{eqnarray}
The outermost bracket $\langle \rangle$ means gauge-configuration average. 

Since we do not so far specify 
whether the gauge-configuration has include the effect of the dynamical quark or not, 
the relation (\ref{RelOrig}) is valid in full QCD and at the quenched level. 
Similarly, the relation is satisfied in finite temperature and density, 
and in confinement and deconfinement phases, and in chiral broken and restored phases. 
Of course, 
by the effects of the dynamical quarks, 
various quantities can change, for example 
Polyakov loop $L_P$, the Dirac eigenvalue distribution $\rho(\lambda)$, 
and the matrix elements $\langle n|\hat{U}_\mu| m \rangle$. 
However, the relation Eq.(\ref{RelOrig}) is satisfied even in full QCD. 

\subsubsection{Discussion for relation between confinement and chiral symmetry breaking} 
Since the relation (\ref{RelOrig}) is the Dirac spectrum representation of the Polyakov loop, 
we can investigate each Dirac mode contribution to the Polyakov loop individually. 
Because the Polyakov loop and Dirac modes 
are important for quark confinement and chiral symmetry breaking, respectively, 
we can discuss the relation between confinement 
and chiral symmetry breaking in QCD using the relation (\ref{RelOrig}). 

The Dirac matrix element of the link-variable operator 
$\langle n|\hat{U}_4| n \rangle$ is generally nonzero. 
In fact, we numerically confirm that and show later. 
Because of the damping factor $\lambda_n^{N_4-1}$, 
the contribution $\lambda_n^{N_4-1}\langle n|\hat{U}_\mu| m \rangle$
from low-lying Dirac modes with $|\lambda_n|\simeq 0$ 
is negligibly small compared to the other Dirac-mode contribution. 
If the matrix element $\langle n|\hat{U}_4| n \rangle$ is large in the low-lying Dirac modes 
stronger than $1/\lambda_n^{N_4-1}$, 
the low-lying Dirac modes have large contribution to the Polyakov loop. 
However, 
even if the behavior of the matrix element $\langle n|\hat{U}_4| n \rangle$ is $\delta$-function  $\delta(\lambda_n)$, 
the contribution from low-lying Dirac modes $\lambda_n^{N_4-1}\delta(\lambda_n)$ is still negligible 
because of $\lambda_n \delta(\lambda_n)=0.$ 
In fact, our all the numerical results show that the low-lying Dirac modes have little contribution \cite{DSIPRD}. 

While the low-lying Dirac modes are essential for chiral symmetry breaking, 
these modes are not essential for confinement. 
In other words, essential modes for confinement and chiral symmetry breaking are different. 
This is consistent with the previous numerical 
lattice result that confinement properties, such as interquark potential and the Polyakov loop, 
are almost unchanged by removing low-lying Dirac modes from the QCD vacuum \cite{GIS}. 
Moreover, even if the chiral restoration occurs and low-lying Dirac modes disappear, 
the contribution from low-lying Dirac modes to the Polyakov loop does not changed 
because the contribution is originally negligible. 
These results suggest that the relation between confinement and chiral symmetry breaking 
is not direct one-to-one corresponding. 

Note that, while the Polyakov loop is defined by gauge fields alone, 
it has a connection to the Dirac modes via the relation (\ref{RelOrig}). 
For example, although the instantons are also defined by gauge fields alone, 
there is a relation between the instantons and the axial U(1) anomaly, 
which is related to a fermionic symmetry. 
Thus, it is not unnatural that the Polyakov loop has a connection to the Dirac modes, 
which are also fermionic modes. 

\section{Modified Kogut-Susskind formalism for temporally odd-number lattice}
In this section, we explain the Modified Kogut-Susskind (KS) formalism, a method for spin-diagonalizing 
the Dirac operator on the temporally odd-number lattice \cite{DSIPRD,DSI}. 
The usual KS formalism \cite{KS} is applicable to only the even lattice, 
where all the lattice sizes are even number with 
periodic boundary condition for link-variables. 
However, the modified KS formalism is applicable to also the temporally odd number lattice, 
where the temporal lattice size is odd number with periodic boundary condition. 

In the modified KS formalism, a matrix $M(s)$ is defined as 
\begin{align}
M(s)\equiv\gamma_1^{s_1}\gamma_2^{s_2}\gamma_3^{s_3}\gamma_4^{s_1+s_2+s_3}, \label{M}
\end{align}
and it is independent of the time component of the site $s_4$.  
Using the matrix $M(s)$, all the $\gamma-$matrices are transformed to be proportional to $\gamma_4$: 
\begin{align}
M^\dagger(s)\gamma_\mu M(s\pm\hat{\mu})=\eta_\mu(s)\gamma_4, \label{MgammaM}
\end{align}
where staggered phase $\eta_\mu(s)$ is defined as 
\begin{align}
\eta_1(s)\equiv 1, \ \ \eta_\mu(s)\equiv (-1)^{s_1+\cdots+s_{\mu-1}} 
\ (\mu \geq 2). \label{eta}
\end{align}
In the Dirac representation, $\gamma_4$ is diagonal as 
$ \gamma_4={\rm diag}(1,1,-1,-1)$, 
and we take the Dirac representation in this paper. 
Thus, the Dirac operator $\slashbb{D}=\gamma_\mu D_\mu$ is spin-diagonalized: 
\begin{align}
\sum_\mu M^\dagger(s) \gamma_\mu D_\mu M(s+ \hat \mu) 
= {\rm diag}(\eta_\mu D_\mu,\eta_\mu D_\mu,-\eta_\mu D_\mu,-\eta_\mu D_\mu), \label{MDiracM}
\end{align}
where the KS Dirac operator $\eta_\mu D_\mu$ is defined as 
\begin{align}
(\eta_\mu D_\mu)_{ss'}=\frac{1}{2a}\sum_{\mu=1}^{4}\eta_\mu(s)\left[U_\mu(s)\delta_{s+\hat{\mu},s'}-U_{-\mu}(s)\delta_{s-\hat{\mu},s'}\right]. \label{KSDiracOp}
\end{align}

Note that the modified KS formalism is applicable to the temporally odd number lattice 
with the periodic boundary condition 
since the periodic boundary condition for the matrix $M(s)$ 
\begin{align}
M(s+N_\mu\hat{\mu})=M(s) \ \ \ \ \ \ \ (\mu=1,2,3,4). \label{OddPBC}
\end{align}
is satisfied. 
Moreover, the periodic boundary condition for the staggered phase $\eta_\mu(s)$ 
is also satisfied on the temporally odd-number lattice. 

Equation (\ref{MDiracM}) is important for reducing the numerical costs. 
Since there are only $\pm\eta_\mu D_\mu$ in Eq. (\ref{MDiracM}), 
all the Dirac eigenvalues $i\lambda_n$ are obtained 
by solving the KS Dirac eigenvalue equation 
\begin{align}
\eta_\mu D_\mu|n) =i\lambda_n|n ) \label{KSEigenEq}
\end{align}
with the KS Dirac eigenstate $|n)$. 
The KS eigenvalue equation is explicitly expressed as 
\begin{align}
\frac{1}{2a}\sum_{\mu=1}^4 
\eta_\mu(s)[U_\mu(s) \chi_n(s+\hat \mu)-U_{-\mu}(s)
\chi_n(s-\hat \mu)] =i\lambda_n\chi_n(s). \label{KSEigenExp}
\end{align}

Moreover, 
the Dirac matrix elements of link-variable operator $\langle n|\hat{U}_4| n \rangle$ 
can be expressed in terms of 
the KS Dirac matrix elements of link-variable operator $(n|\hat{U}_4|n)$: 
\begin{align}
\langle n|\hat{U}_4| n \rangle=(n|\hat{U}_4| n). \label{U4DiracKSOddDiag}
\end{align}
Because of the degeneracy of the Dirac eigenvalues $i\lambda_n$ and Eq.(\ref{U4DiracKSOddDiag}), 
the analytical relation (\ref{RelOrig}) can be rewritten as 
\begin{align}
 L_P =\frac{(2ai)^{N_4-1}}{3V}
\sum_{n}\lambda_n^{N_4-1}(n|\hat{U}_4|n) \label{RelKS}
\end{align}
using the modified KS formalism. 

Note that the (modified) KS formalism is an exact mathematical 
method for diagonalizing the Dirac operator and is not an approximation. 
Thus, Eqs. (\ref{RelOrig}) and (\ref{RelKS}) are completely equivalent. 
While the dimension of the Dirac operator $\slashb{D}$ is $(4\times N_c\times V)^2$, 
that of the KS Dirac operator $\eta_\mu D_\mu$ is $(N_c\times V)^2$. 
Thus, the numerical costs can be reduced by using the modified KS formalism. 
In our study, we just use the KS formalism as the technique to reduce the numerical costs, 
and we do not use a specific fermion like the KS fermion. 
Actually, even if we do not use the KS formalism, 
we can obtain the same results. 
However, the numerical cost is simply larger. 

\section{Lattice QCD numerical analysis}
As we discuss in the previous section, 
the analytical relation (\ref{RelOrig}) indicates small contribution from low-lying Dirac modes 
to the Polyakov loop. 
However, it is also important to confirm that quantitatively. 
In particular, 
it is worth investigating the properties of 
the (KS) Dirac matrix elements of link variable operators $(n|\hat{U}_4| n)$. 
In this section, we numerically perform SU(3) lattice QCD calculations and 
quantitatively discuss the relation between confinement and chiral symmetry breaking 
based on the relation (\ref{RelKS}) connecting the Polyakov loop 
and Dirac modes on the temporally-odd number lattice. 

The SU(3) lattice QCD Monte Carlo simulations are performed 
with the standard plaquette action at the quenched level 
in both cases of confinement and deconfinement phases. 
We use a $10^3\times5$ lattice, and two values for $\beta\equiv\frac{2N_{\rm c}}{g^2}$. 
The confinement phase is produced with $\beta=5.6$ (i.e., $a\simeq0.25$ fm), 
corresponding to $T\equiv1/(N_4 a)\simeq160$ MeV, 
and the deconfinement phase is produced with $\beta=6.0$ (i.e., $a\simeq0.10$ fm), 
corresponding to $T\equiv1/(N_4 a)\simeq400$ MeV. 
For each phase, we use 20 gauge configurations, 
which are taken every 500 sweeps after the thermalization of 5,000 sweeps. 

\subsection{Small contribution of low-lying Dirac modes to Polyakov loop}
Using the relation (\ref{RelKS}), we numerically calculate the each Dirac mode contribution 
$\frac{(2ai)^{N_4-1}}{3V}\lambda_n^{N_4-1}(n|\hat{U}_4| n)$ to the Polyakov loop. 
In particular, since RHS of Eq. (\ref{RelKS}) is expressed 
as a sum of the Dirac-mode contribution, 
we can calculate the Polyakov loop without low-lying Dirac-mode contribution as 
\begin{align}
(L_P)_{\rm IR\hbox{-}cut}=\frac{(2ai)^{N_4-1}}{3V}
\sum_{|\lambda_n|>\Lambda_{\rm IR}}\lambda_n^{N_4-1}(n|\hat{U}_4|n), \label{IRCutPolyakov}
\end{align}
with the infrared (IR) cutoff $\Lambda_{\rm IR}$ for Dirac eigenvalue. 
The chiral condensate without the low-lying Dirac mode contribution 
under IR cutoff $\Lambda_{\rm IR}$ is also expressed as 
\begin{align}
\langle \bar{q}q\rangle_{\Lambda_{\rm IR}}
=-\frac{1}{V}\sum_{\lambda_n\geq\Lambda_{\rm IR}}\frac{2m}{\lambda_n^2+m^2}
\end{align}
In the chiral broken phase, 
using the IR cutoff $\Lambda_{\rm IR}\simeq0.4 {\rm GeV}$ 
and physical current-quark mass, $m\simeq5 {\rm MeV}$, 
the chiral condensate is largely reduced to about 2\% as 
$\frac{\langle \bar{q}q\rangle_{\Lambda_{\rm IR}}}{\langle \bar{q}q\rangle} \simeq 0.02$ 
and chiral symmetry is almost restored \cite{GIS}. 
In this paper, we take the IR cutoff of $\Lambda_{\rm IR}\simeq0.4 {\rm GeV}$.

We numerically confirm that the low-lying Dirac modes have little contribution 
in both confinement and deconfinement phases. 
In the deconfinement phase, the $Z_3$ center symmetry is spontaneously broken, 
and the Polyakov loop is proportional to 
$\mathrm{e}^{i\frac{2\pi}{3}j} \ (j=0,\pm 1)$ 
for each gauge configuration at the quenched level \cite{Rothe}. 
In this paper, we name the vacuum where the Polyakov loop is almost real 
($j$=0) ``real Polyakov-loop vacuum'' 
and the other vacua ``$Z_3$-rotated vacua.'' 
At the quenched level, 
these three vacua are degenerated vacua. 
In full QCD, the degeneracy is resolved, 
and the real Polyakov-loop vacuum is selected as the stable vacuum 
while the $Z_3$-rotated vacua become metastable states. 
The numerical results for the $Z_3$-rotated vacua in the deconfinement phase are discussed 
in Ref. \cite{DSIPRD}. 
In this paper, 
we concentrate the results on the real Polyakov-loop vacuum in the deconfinement phase. 

In Tables \ref{LowConf} and \ref{LowDeconf}, 
the numerical results for $L_P$ and $(L_P)_{\rm IR\hbox{-}cut}$ are shown
in the confinement and deconfinement phases. 
From Table \ref{LowConf} and \ref{LowDeconf}, 
the relation $L_P\simeq (L_P)_{\rm IR\hbox{-}cut}$ 
is almost satisfied for each gauge-configuration in both confinement and deconfinement phases. 
Therefore, the low-lying Dirac modes have little contribution to the Polyakov loop 
and are not essential for confinement. 
However, the low-lying Dirac modes 
below the IR cutoff $|\lambda_n|<\Lambda_{\rm IR}\simeq0.4 {\rm GeV}$ are essential 
for chiral symmetry breaking. 
Thus, it is suggested that 
there is no direct one-to-one correspondence between confinement and chiral symmetry breaking. 

\begin{table}[htb]
\caption{
Numerical results for $L_P$ and 
$(L_P)_{\rm IR\hbox{-}cut}$ 
in lattice QCD with $10^3\times5$ and $\beta=5.6$ 
for each gauge configuration, 
where the system is in the confinement phase.
}
  \begin{tabular}{ccccccccc} \hline 
Configuration No. &1&2&3&4&5&6&7&8 \\ \hline
Re$L_P$
&0.00961 &-0.00161&0.0139    &-0.00324&0.000689
&0.00423 &-0.00807&-0.00918\\ 
Im$L_P$             
&-0.00322&-0.00125&-0.00438&-0.00519&-0.0101  
&-0.0168 &-0.00265&-0.00683\\
Re$(L_P)_{\rm IR\hbox{-}cut}$ 
&0.00961 &-0.00160&0.0139    &-0.00325&0.000706
&0.00422&-0.00807 &-0.00918\\ 
Im$(L_P)_{\rm IR\hbox{-}cut}$ 
&-0.00321&-0.00125&-0.00437&-0.00520&-0.0101  
&-0.0168&-0.00264  &-0.00682\\ \hline
  \end{tabular}
\label{LowConf}
\end{table}
\begin{table}[htb]
\caption{
Numerical results for $L_P$ and 
$(L_P)_{\rm IR\hbox{-}cut}$ 
in lattice QCD with $10^3\times5$ and $\beta=6.0$ 
for each gauge configuration, 
where the system is in the deconfinement phase and the real Polyakov-loop vacuum.
}
  \begin{tabular}{ccccccccc} \hline 
Configuration No. &1&2&3&4&5&6&7&8 \\ \hline
Re$L_P$             
&0.316     &0.337       &0.331     &0.305     &0.314   
&0.316     &0.337       &0.300      \\
Im$L_P$              
&-0.00104&-0.00597  &0.00723  &-0.00334&0.00167
&0.000120&0.0000482&-0.00690\\
Re$(L_P)_{\rm IR\hbox{-}cut}$ 
&0.319     &0.340       &0.334     &0.307     &0.317   
&0.319      &0.340      &0.303      \\
Im$(L_P)_{\rm IR\hbox{-}cut}$ 
&-0.00103&-0.00597  &0.00724  &-0.00333&0.00167
&0.000121 &0.0000475&-0.000691\\ \hline
  \end{tabular}
\label{LowDeconf}
\end{table}

\subsection{Properties of Dirac-mode matrix element of link-variable operator}
Next, we quantitatively investigate 
the properties of the each Dirac-mode contribution to the Polyakov loop 
$\lambda_n^{N_4-1}(n|\hat{U}_4|n)$ and 
in particular the (KS) Dirac-mode matrix elements of the link-variable operators 
$(n|\hat{U}_4| n)$ in both confinement and deconfinement phases. 

\subsubsection{Confinement phase}
\begin{figure}[h]
%\begin{center}
\includegraphics[scale=0.5]{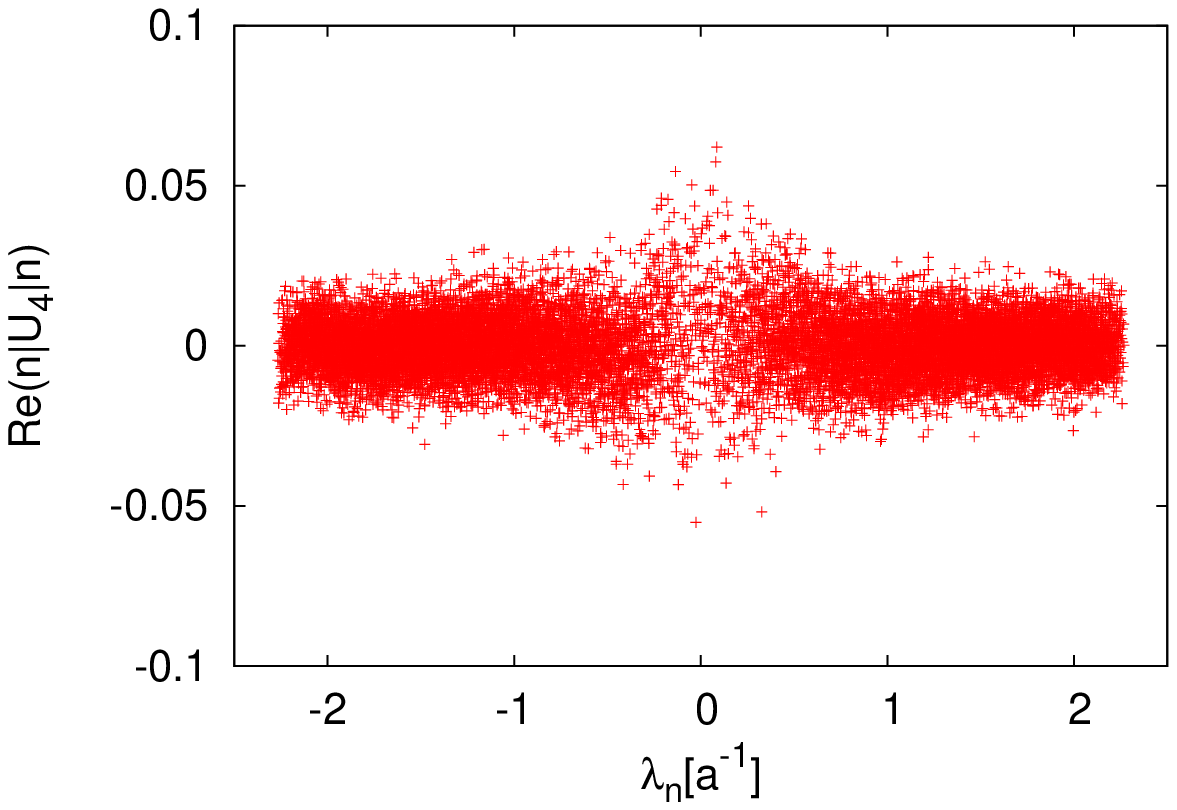}
\includegraphics[scale=0.5]{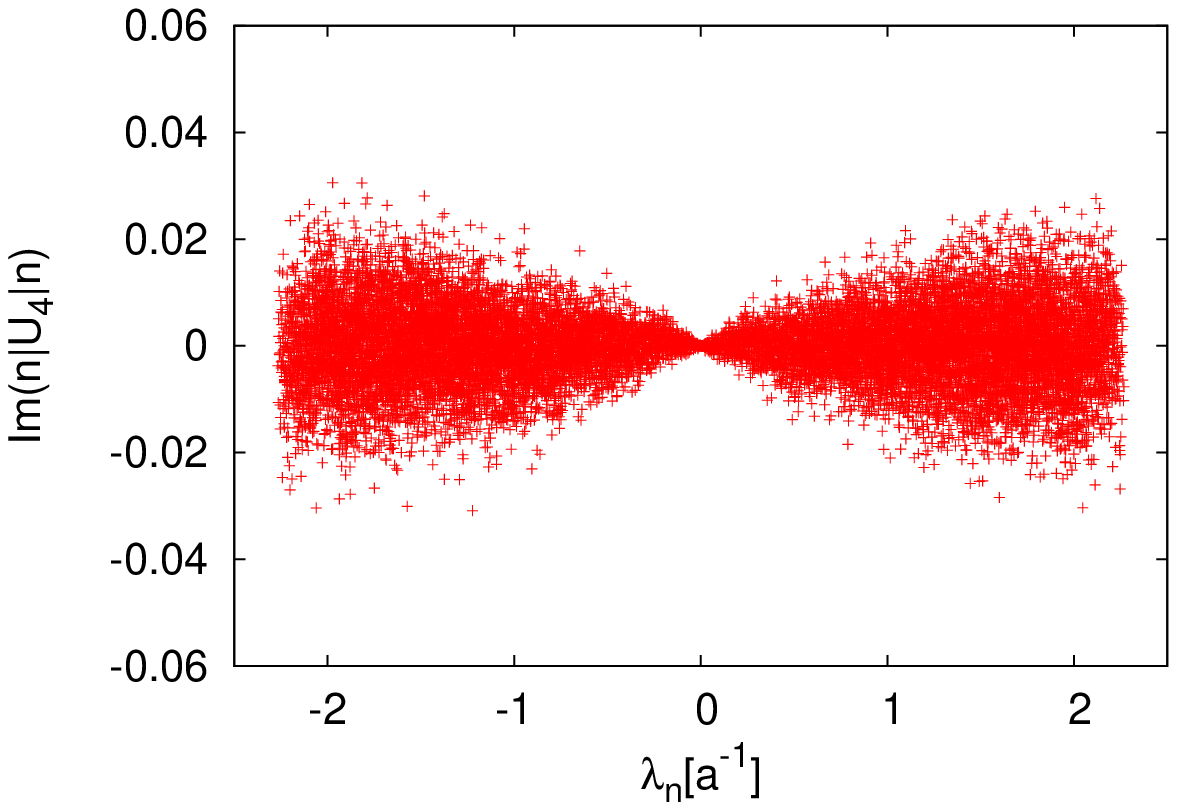}
\caption{
The numerical results for the real part Re$(n|\hat{U}_4|n)$ 
and the imaginary part Im$(n|\hat{U}_4|n)$ of the matrix element 
in the confinement phase, 
plotted against the Dirac eigenvalue $\lambda_n$ 
in the lattice unit at $\beta=5.6$ on $10^3\times 5$ 
taken from Ref. \cite{DSIPRD}. 
There is the positive/negative symmetry. 
}
\label{MatEleConf}
%\end{center}
\end{figure}

Figure \ref{MatEleConf} shows the numerical results for the matrix elements 
Re$(n|\hat{U}_4|n)$ and Im$(n|\hat{U}_4|n)$ 
plotted against Dirac eigenvalues $\lambda_n$ in the lattice unit 
for one gauge configuration in the confinement phase. 

\begin{figure}[h]
%\begin{center}
\includegraphics[scale=0.5]{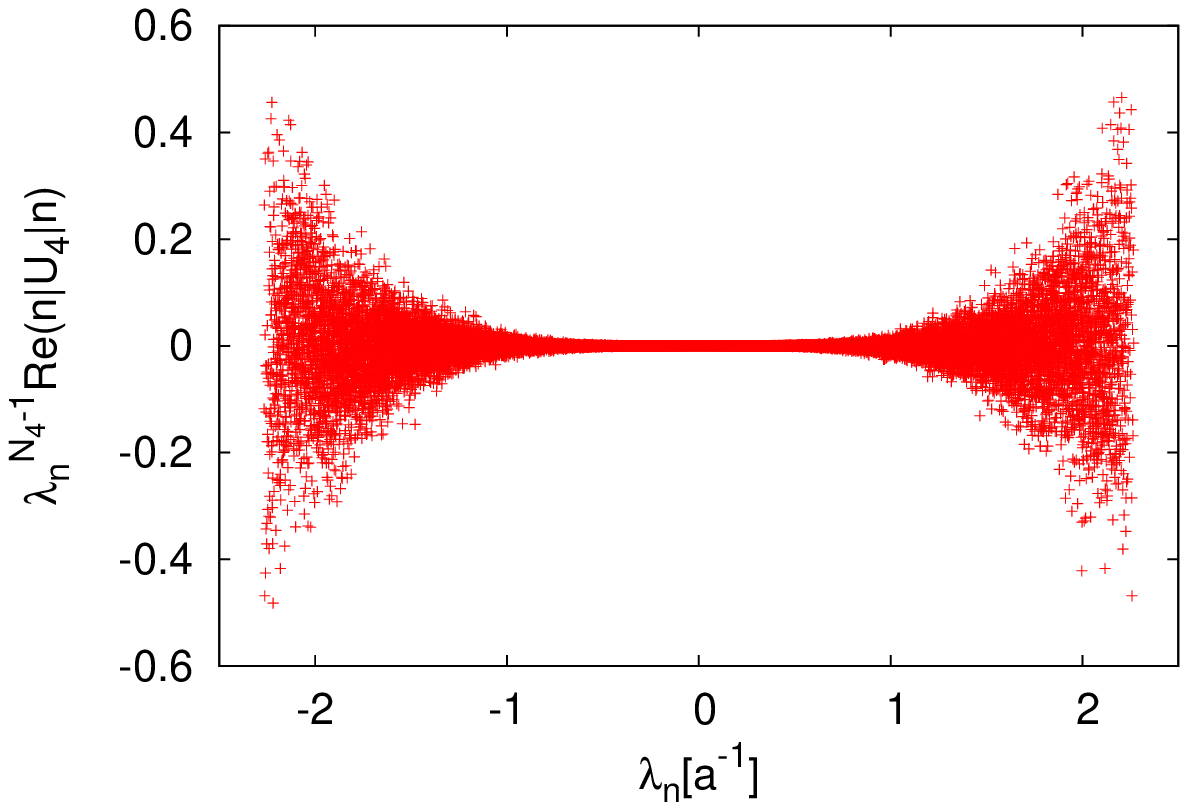}
\includegraphics[scale=0.5]{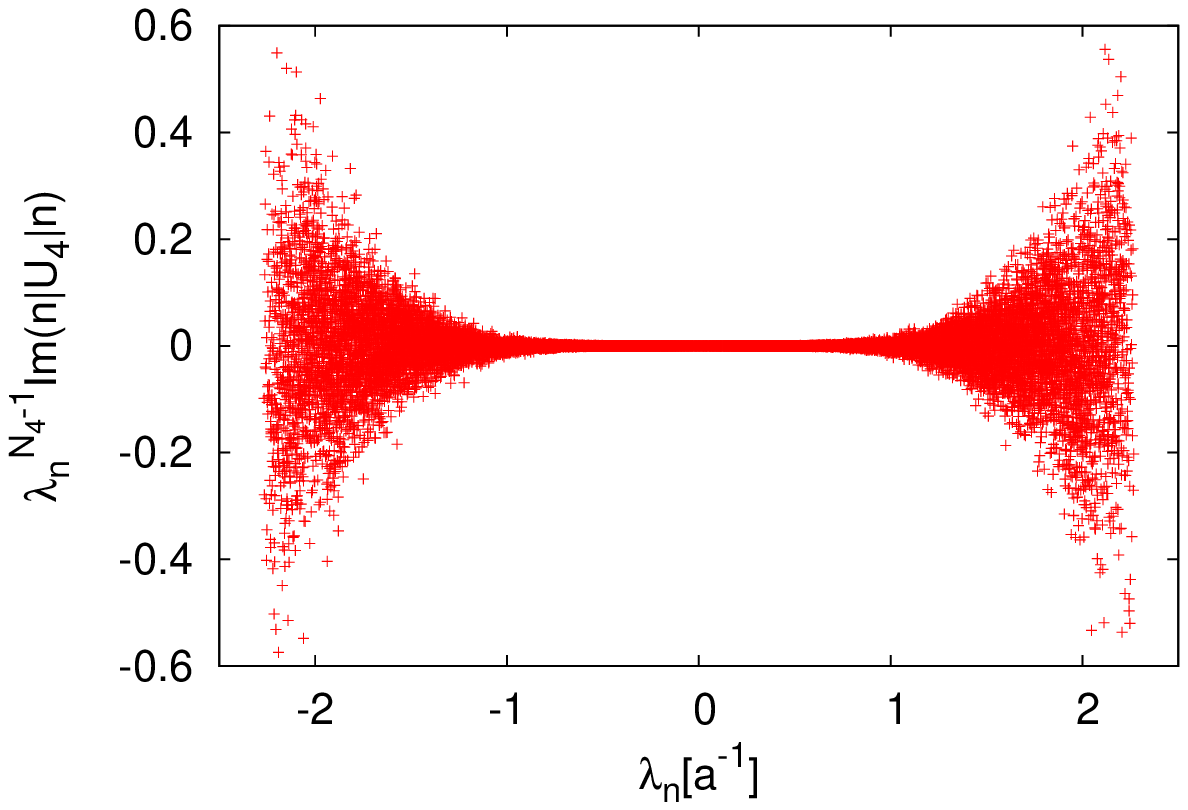}
\caption{
The numerical results for each Dirac-mode contribution to the Polyakov loop, 
$\lambda_n^{N_4-1}{\rm Re}(n|\hat{U}_4|n)$ 
and $\lambda_n^{N_4-1}{\rm Im}(n|\hat{U}_4|n)$ 
in the confinement phase, 
plotted against the Dirac eigenvalue $\lambda_n$ 
in the lattice unit 
at $\beta=5.6$ on $10^3\times5$ 
taken from Ref. \cite{DSIPRD}. 
There is the positive/negative symmetry and it lead to $L_P=0$ in the confinement phase. 
}
\label{ContConf}
%\end{center}
\end{figure}

Figure~\ref{ContConf} shows each Dirac-mode contribution to the Polyakov loop 
$\lambda_n^{N_4-1}{\rm Re}(n|\hat{U}_4|n)$ and $\lambda_n^{N_4-1}{\rm Im}(n|\hat{U}_4|n)$ 
plotted against Dirac eigenvalues $\lambda_n$ in the lattice unit. 

While the real part of the matrix element ${\rm Re}(n|\hat{U}_4|n)$ 
is large in low-lying Dirac-mode region from Fig. \ref{MatEleConf}, 
the contribution from the low-lying Dirac modes to the Polyakov loop, 
$\lambda_n^{N_4-1}{\rm Re}(n|\hat{U}_4|n)$, 
is small and negligible because of the damping factor $\lambda_n^{N_4-1}$ 
from Fig. \ref{ContConf}. 
Thus, as we expect, the damping factor $\lambda_n^{N_4-1}$ has an essential role in 
Eq.(\ref{RelKS}). 

On the other hand, from Fig. \ref{MatEleConf}, 
the imaginary part ${\rm Im}(n|\hat{U}_4|n)$ of the matrix element 
is relatively small in low-lying Dirac-mode region than that in other region, 
unlike the real part ${\rm Re}(n|\hat{U}_4|n)$. 
In any case, $\lambda_n^{N_4-1}{\rm Im}(n|\hat{U}_4|n)$ 
is small in low-lying Dirac-mode region, 
as shown in Fig. \ref{ContConf}. 

An important point is that 
the distribution of Dirac-mode matrix element $(n|\hat{U}_4|n)$, 
i.e., ${\rm Re}(n|\hat{U}_4|n)$ and ${\rm Im}(n|\hat{U}_4|n)$, 
has the symmetry on the positive and negative values in the whole Dirac-mode region 
in the confinement phase from Fig. \ref{MatEleConf}. 
We named the symmetry ``positive/negative symmetry'' \cite{DSIPRD,DSI}. 
Then, 
the distribution of each Dirac-mode contribution to the Polyakov loop, $\lambda_n^{N_4-1}(n|\hat{U}_4|n)$,  has the same symmetry in the confinement phase. 
This symmetry leads to the relation 
\begin{eqnarray}
\sum_{\Lambda_1\leq\lambda_n\leq\Lambda_2}\lambda_n^{N_4-1}(n|\hat{U}_4| n)
=0 
\ \ ({\rm confinement \ phase}) \label{pnsym}
\end{eqnarray}
with arbitrary $\Lambda_1$ and $\Lambda_2$. 
The relation (\ref{pnsym}) means that 
the contribution from arbitrary Dirac-mode region to the Polyakov loop is zero. 
This behavior in the confinement phase is consistent with the previous works \cite{GIS}. 
Therefore, because of the positive/negative symmetry, the Polyakov loop is zero, i.e., $\langle L_P\rangle=0$,  in the confinement phase. 
Note that the distribution of 
the matrix elements $(n|\hat{U}_4|n)$ 
is not statistical fluctuation on the gauge ensemble 
because the results shown here are for one configuration. 
We confirm the same behavior for other gauge configurations. 

As for the $N_4$ dependence of the matrix element $(n|\hat{U}_4|n)$ in the confinement phase, 
we find almost the same results that 
there is the positive/negative symmetry 
and low-lying Dirac modes have little contribution to the Polyakov loop. 

\subsubsection{Deconfinement phase case}

Next, we numerically investigate the Dirac-mode matrix element $(n|\hat{U}_4|n)$ and 
each Dirac-mode contribution to the Polyakov loop $\lambda_n^{N_4-1}(n|\hat{U}_4|n)$ 
in the deconfinement phase. 
Since the deconfinement phase does not have confinement and chiral symmetry breaking, 
it may be less interesting to consider their relation there. 
While 
there are the real Polyakov-loop vacuum and two $Z_3$-rotated vacua in the deconfinement phase, 
we consider only the real Polyakov-loop vacuum 
because it is selected as the stable vacuum in full QCD. 

\begin{figure}[h]
%\begin{center}
\includegraphics[scale=0.5]{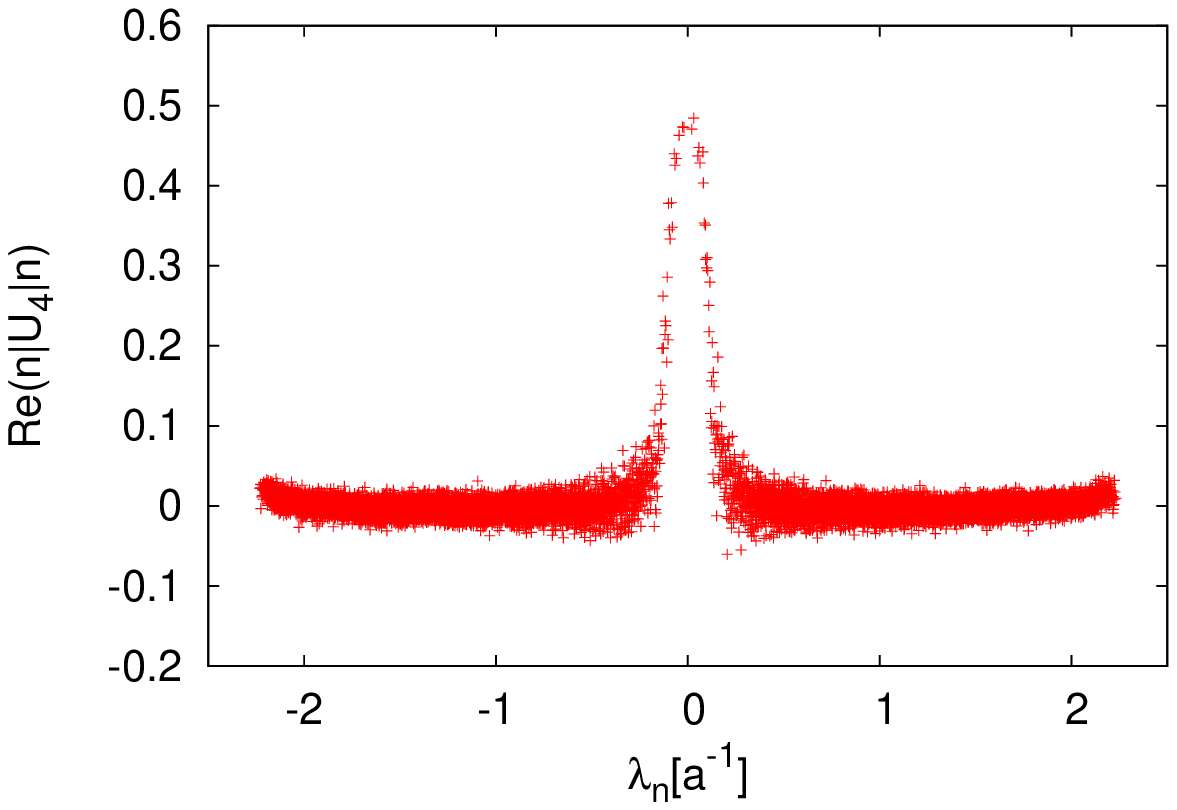}
\includegraphics[scale=0.5]{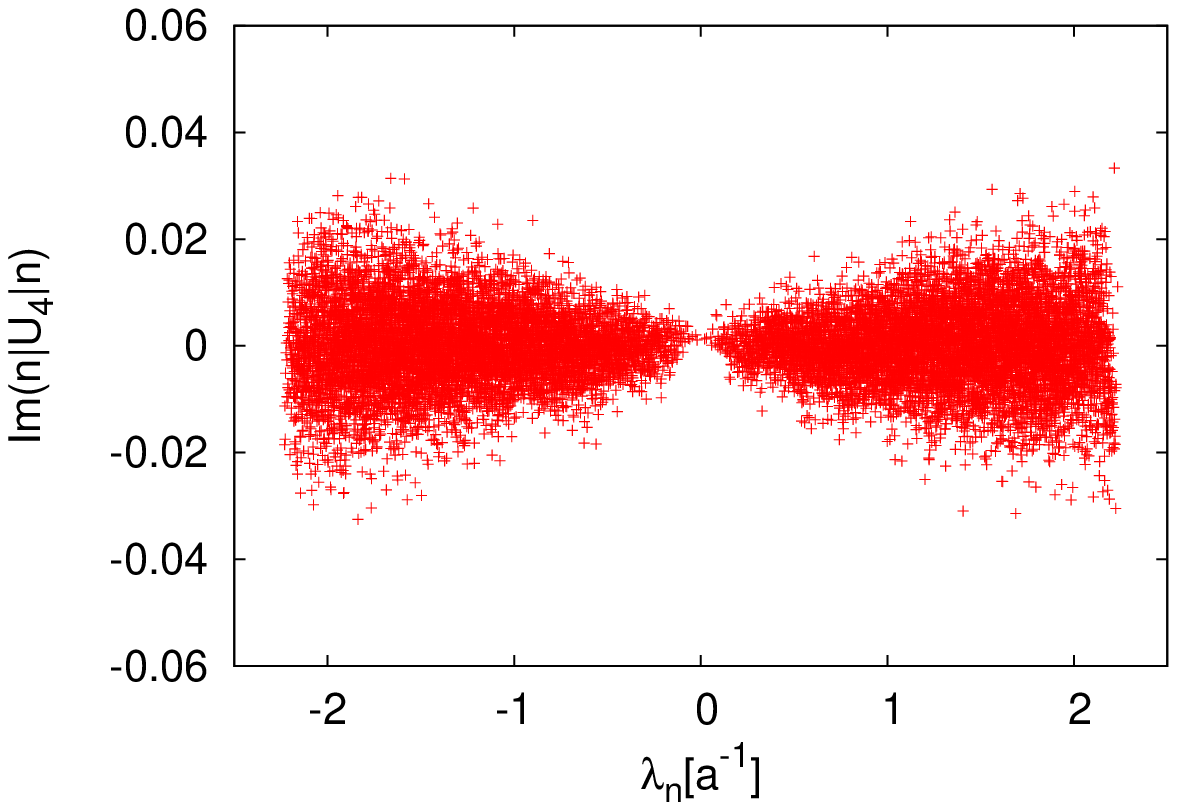}
\caption{
The numerical results for the real part Re$(n|\hat{U}_4|n)$ 
and the imaginary part Im$(n|\hat{U}_4|n)$ of the matrix element 
in the deconfinement phase and the real Polyakov-loop vacuum, 
plotted against the Dirac eigenvalue $\lambda_n$
in the lattice unit 
at $\beta=6.0$ on $10^3\times5$. 
The real part is taken from Ref. \cite{DSIPRD}. 
}
\label{MatEleDeconf}
%\end{center}
\end{figure}

\begin{figure}[h]
%\begin{center}
\includegraphics[scale=0.5]{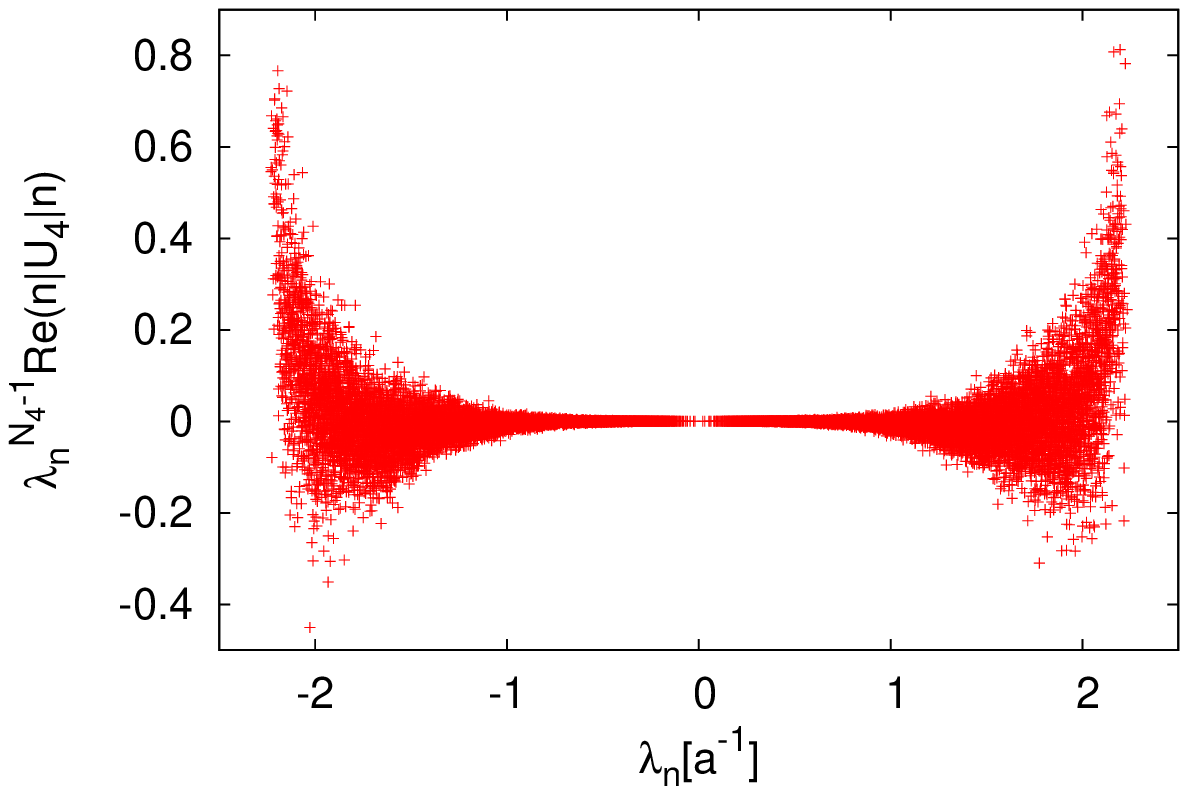}
\includegraphics[scale=0.5]{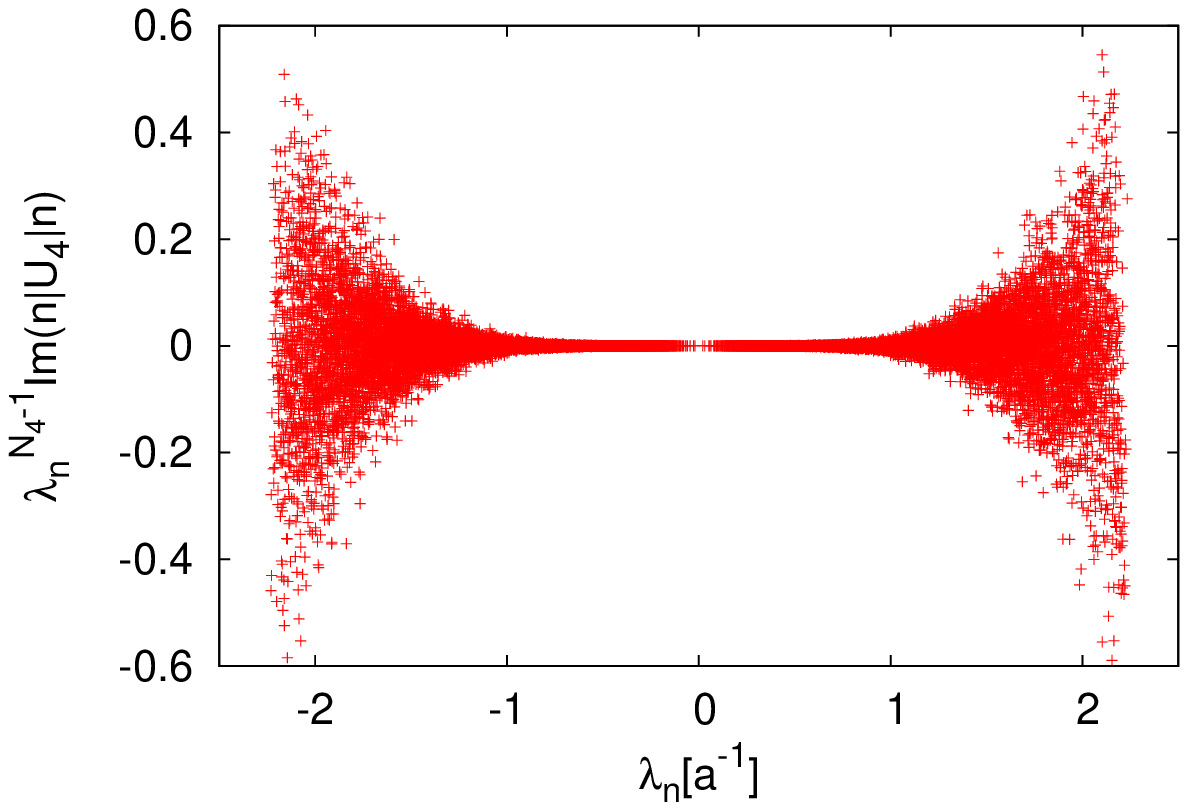}
\caption{
The numerical results for each Dirac-mode contribution to the Polyakov loop, 
$\lambda_n^{N_4-1}{\rm Re}(n|\hat{U}_4|n)$ and 
$\lambda_n^{N_4-1}{\rm Im}(n|\hat{U}_4|n)$, 
in the deconfinement phase and the real Polyakov-loop vacuum, 
plotted against the Dirac eigenvalue $\lambda_n$ in the lattice unit 
at $\beta=6.0$ on $10^3\times5$. 
The real part is taken from Ref. \cite{DSIPRD}. 
}
\label{ContDeconf}
%\end{center}
\end{figure}

We show in Figs. \ref{MatEleDeconf} and \ref{ContDeconf} 
the matrix elements and each Dirac-mode contribution 
in the deconfinement phase and the real Polyakov-loop vacuum, 
plotted against the Dirac eigenvalue $\lambda_n$, in quenched lattice QCD. 
The imaginary parts of the matrix element, ${\rm Im}(n|\hat{U}_4|n)$, 
and the each Dirac-mode contribution, $\lambda_n^{N_4-1}{\rm Im}(n|\hat{U}_4|n)$,
show the same behavior as the case of the confinement phase 
because the imaginary part of the Polyakov loop is zero in the real-Polyakov loop vacuum. 
(Compare Fig. \ref{MatEleConf}(b) and Fig. \ref{MatEleDeconf}(b).) 
Then, we consider only the results for real part of these quantities 
in the deconfinement phase. 
Like the case of the confinement phase, 
we show the results for one gauge configuration. 

There is a peak low-lying Dirac-mode region 
of the real part of the matrix element, ${\rm Re}(n|\hat{U}_4|n)$, 
from Fig. \ref{MatEleDeconf}. 
However, from Fig. \ref{ContDeconf}, each Dirac-mode contribution 
$\lambda_n^{N_4-1}{\rm Re}(n|\hat{U}_4|n)$ is 
relatively small in low-lying Dirac-mode region because of the damping factor $\lambda_n^{N_4-1}$ 
like the case of confinement phase. 
More quantitatively, 
only high-lying Dirac modes have contribution to the nonzero value of the Polyakov loop 
from Fig. \ref{ContDeconf}. 
In the deconfinement phase, there is no positive/negative symmetry for 
the distributions of the matrix element ${\rm Re}(n|\hat{U}_4|n)$ 
and each Dirac-mode contribution $\lambda_n^{N_4-1}{\rm Re}(n|\hat{U}_4|n)$, 
unlike the case of the confinement phase. 
The Polyakov loop is nonzero 
because of the asymmetry in the distribution of the matrix element and each Dirac-mode contribution, 
while the Polyakov loop in the confinement phase is zero because of the symmetry. 
Thus, the behavior of the matrix element $(n|\hat{U}_4|n)$ on the positive/negative symmetry 
is strongly related to the deconfinement phase transition. 
The positive/negative symmetry and the $Z_3$ center symmetry 
have common point that 
are not broken in the confinement phase and are broken in the deconfinement phase 
at the quenched level. 
Therefore, it is interesting to investigate the relation between the new positive/negative symmetry 
and the $Z_3$ center symmetry. 

Although we consider only the real Polyakov-loop vacuum in this paper, 
we considered $N_4$ dependence of the matrix element $(n|\hat{U}_4|n)$ 
in the deconfinement phase and the behavior of the matrix element in $Z_3$ rotated vacua 
in Ref. \cite{DSIPRD}. 
However, qualitative behaviors are same as the results for the results shown here. 

\section{Summary and concluding remarks}
We have studied the relation between confinement and chiral symmetry breaking in QCD 
using the Dirac spectrum representation of the Polyakov loop (\ref{RelOrig}) 
based on the lattice QCD formalism. 
In this paper, we have derived the analytical relation (\ref{RelOrig}) 
on the temporally odd-number lattice with the normal periodic boundary condition for link-variables. 
Since the Polyakov loop is an order parameter of quark confinement 
and Dirac modes are strongly correlated to chiral symmetry breaking in QCD, 
the relation (\ref{RelOrig}) can be used to discuss the relation between these 
non-perturbative phenomena. 

First, we have discussed the properties of the relation (\ref{RelOrig}). 
The relation can be derived in arbitrary gauge theories using few assumptions. 
Odd parity of the temporal lattice size is essentially not required for the derivation \cite{DSIPRD}. 
Moreover, this relation is valid not only at the quenched level also but in the full QCD and in finite  temperature/density. 
The relation indicates negligible contribution to the Polyakov loop from the low-lying Dirac modes 
because of the damping factor $\lambda_n^{N_4-1}$. 
Thus, it is suggested that there is no one-to-one corresponding between 
quark confinement and chiral symmetry breaking in QCD. 

Next, we have discussed the modified KS formalism on the temporally odd-number lattice. 
When solving the Dirac eigenvalue equation, 
the numerical costs are large because of the huge dimension of the Dirac operator $\slashb{D}$. 
Using the normal KS formalism on the even lattice, 
where all the lattice sizes are even number, 
the numerical cost can be reduced. 
Since the normal KS formalism is not applicable to the temporally odd-number lattice, 
we have developed the modified KS formalism applicable to the temporally odd-number lattice, 
and we have used it as a method for reducing the numerical costs. 
We have derive the relation (\ref{RelKS}) which is equivalent 
to the original relation (\ref{RelOrig}) using the modified KS formalism. 

Next, we have performed the numerical lattice QCD Monte Carlo calculation 
with the standard plaquette action at the quenched level in both confinement and deconfinement phases. 
We impose the temporal periodic boundary condition to the temporally odd-number lattice, 
which is required for the imaginary-time formalism at finite temperature. 
It is numerically confirmed that 
the low-lying Dirac modes have little contribution to the Polyakov loop 
in both confinement and deconfinement phases, 
and the damping factor $\lambda_n^{N_4-1}$ in the relation (\ref{RelOrig}) plays an important role. 
Thus, no one-to-one corresponding between confinement and chiral symmetry breaking in QCD 
has been supported by our numerical analysis. 

Also, we have investigated the properties 
of the Dirac-mode matrix element $(n|\hat{U}_4|n)$ 
which appears in the relation (\ref{RelKS}). 
Remarkably, in the confinement phase, there is the positive/negative symmetry 
in the distribution of the matrix element $(n|\hat{U}_4|n)$, 
and thus the Polyakov loop is zero. 
In the deconfinement phase, however, 
the positive/negative symmetry disappears in the distribution of the matrix element $(n|\hat{U}_4|n)$, 
and then the Polyakov loop is nonzero. 
This symmetry distinguishes the confinement and deconfinement phases at the quenched level. 
Since the behavior of the positive/negative symmetry is similar to 
the center symmetry in the pure-gauge theory, which is related to confinement \cite{Greensite}, 
it is interesting to investigate the relation between these symmetries. 

In this study, we have performed the lattice QCD calculation at the quenched level. 
However, the full QCD calculation is desired for more quantitative discussion. 
In particular, it is interesting to investigate the properties of the new positive/negative symmetry 
of the matrix element $(n|\hat{U}_4|n)$ in the full QCD calculation. 

Recently, it was pointed out that the fluctuation of the Polyakov loop 
is important for the deconfinement phase transition \cite{Redlich}. 
The renormalization of the Polyakov loop in the physical continuum limit 
is also the outstanding problem. 
However, one can discuss the deconfinement phase transition 
avoiding the uncertainties of renormalization of the Polyakov loop 
by considering the ratio of susceptibility of the Polyakov loop. 
As a next work, 
it is interesting to investigate the relation between the Polyakov loop fluctuation and Dirac modes 
using our scheme in this paper. 

Since the QCD monopole in the maximally Abelian gauge is important for 
non-perturbative phenomena of low-energy QCD, 
such as confinement and chiral symmetry breaking \cite{Miyamura, Woloshyn}, 
it is expected that the QCD monopole without the low-lying Dirac modes is important for only confinement. 
Thus, it is also interesting to study the relation 
between the QCD monopole and low-lying Dirac modes 
by using gauge-invariant Dirac-mode expansion \cite{GIS}. 

Our analysis indicates the suggestion of possible difference 
between confinement and chiral symmetry breaking in QCD. 
Then, a new phase can exist in QCD, for example 
where chiral symmetry is restored but the quark is confined 
\cite{YAoki, GIS, LS11}. 
Not only finite temperature and quark chemical potential, 
but also strong electromagnetic fields 
can change the structure of the QCD vacuum \cite{SuganumaTatsumi}. 
QCD has possibly such a new phase in the strong electromagnetic fields. 

\section*{Acknowledgements}
H. S. is supported in part by the Grant for Scientific Research [(C) No.23540306, E01:21105006] 
from the Ministry of Education, Culture, Sports, Science and Technology (MEXT) of Japan. 
The lattice QCD calculations were performed on the NEC-SX8R 
and NEC-SX9 at Osaka University. 



\bibliographystyle{aipproc}   % if natbib is available

\bibliography{sample}

\IfFileExists{\jobname.bbl}{}
 {\typeout{}
  \typeout{******************************************}
  \typeout{** Please run "bibtex \jobname" to optain}
  \typeout{** the bibliography and then re-run LaTeX}
  \typeout{** twice to fix the references!}
  \typeout{******************************************}
  \typeout{}
 }

%\end{document}

%%%%%%%%%%%%%%%%%%%%%%%%%%%%%%%%%%%%%%%%%%%
%% The following lines show an example how to produce a bibliography
%% without the help of the BibTeX program. This could be used instead
%% of the above.
%%%%%%%%%%%%%%%%%%%%%%%%%%%%%%%%%%%%%%%%%%%

\end{document}